\documentstyle[prl,aps,multicol,psfig]{revtex}

\def\newbaselineskip{\baselineskip}

\ifpreprintsty\else\textheight = 61\baselineskip\fi

\begin{document}

\draft   



\def\figaa{1 }
\def\figbb{2 }

\def\ch{\rho}
\def\sp{\sigma}

\def\eps{\epsilon}

\def\beq{\begin{equation}}
\def\eeq{\end{equation}}
\def\beqn{\begin{eqnarray}}
\def\eeqn{\end{eqnarray}}

\def\al{\alpha}
\def\g{\sigma}
\def\und{\underline}
\def\Lam{\Lambda}
\def\eqref#1{(\ref{#1})}
\def\up{\uparrow}
\def\down{\downarrow}

\def\ggg#1#2#3#4#5{g_{#1}(#2;#3|#4;#5)}

\def\ccd#1#2#3{c^{\dag}_{r_{#1},\al_{#1},#2,#3}}
\def\cc#1#2#3{c_{r_{#1},\al_{#1},#2,#3}}

\long\def\taglia#1{}
\long\def\tagliab#1{}
\long\def\tagliac#1{}

\long\def\tagliad#1{}

\long\def\added#1{#1}

\def\evi{\it}

\newcommand{\cmin}{\lesssim} 
\newcommand{\cmag}{\gtrsim}

\def\pperp{p}
\def\KF{K}

\def\tilU{{\cal U}}

\def\prl{Phys. Rev. Lett. }
\def\prb{Phys. Rev. B }
\def\ut#1{{\bf #1}}

\def\mycaption#1{ {\small #1}}

  \def\btwocols{}
  \def\etwocols{}

\def\makecols{
  \def\btwocols{\begin{multicols}{2}}
  \def\etwocols{\end{multicols}}
}

\ifpreprintsty\else\makecols\fi

\ifpreprintsty
\def\psfig#1{} \def\mycaption#1{{\normalsize #1}} 
\def\efiga{\figa}
\def\efigb{\figb}
\def\ifiga{}
\def\ifigb{}
\else
\def\ifiga{\figa}
\def\ifigb{\figb}
\def\efiga{}
\def\efigb{}
\fi

\title{
Spin and charge excitations in a Three-Legs Fermionic Ladder:
a 
Renormalization-Group study.
}

\author{Enrico Arrigoni 
}
\address{ 
Institut f\"ur Theoretische Physik
Universit\"at W\"urzburg,
D-97074 W\"urzburg, Germany\\
Tel. +49-931-888-5892; Fax. +49  931 888-5141; E-Mail: arrigoni@physik.uni-wuerzburg.de }

\maketitle

\begin{abstract}

We study the spin and charge  phase diagram of a three-legs ladder 
(at zero temperature) 
as a function of  fermion density  and of 
transverse single-particle hopping by means of a 
Renormalization-Group analysis rigorously  controlled in the
weak-coupling limit. 

Periodic boundary conditions in the
 direction transverse to the ladder 
 produce frustrated magnetic excitations yielding a
spin-gapped phase in a large region about and at half filling.
Spin correlations are instead enhanced
 when open
transverse boundary conditions are considered,
yielding an ungapped phase in a wide region
 about half filling
 (as observed in the
Sr$_{n-1}$Cu$_{n+1}$O$_{2n}$ stripes compounds)
and at
half filling down to a critical value of the transverse hopping $t_{\perp}$.
At that critical value, the system 
undergoes a Mott transition at half filling
by decreasing $t_{\perp}$.

\end{abstract}

\pacs{PACS numbers : 72.15.Nj, 64.60.Ak, 71.27.+a}



\baselineskip=\newbaselineskip


\long\def\figa{
\begin{figure}[htb]
\centerline{
\psfig{width=9truecm,angle=-90,bbllx=50pt,bblly=100pt,bburx=560pt,bbury=750pt,file={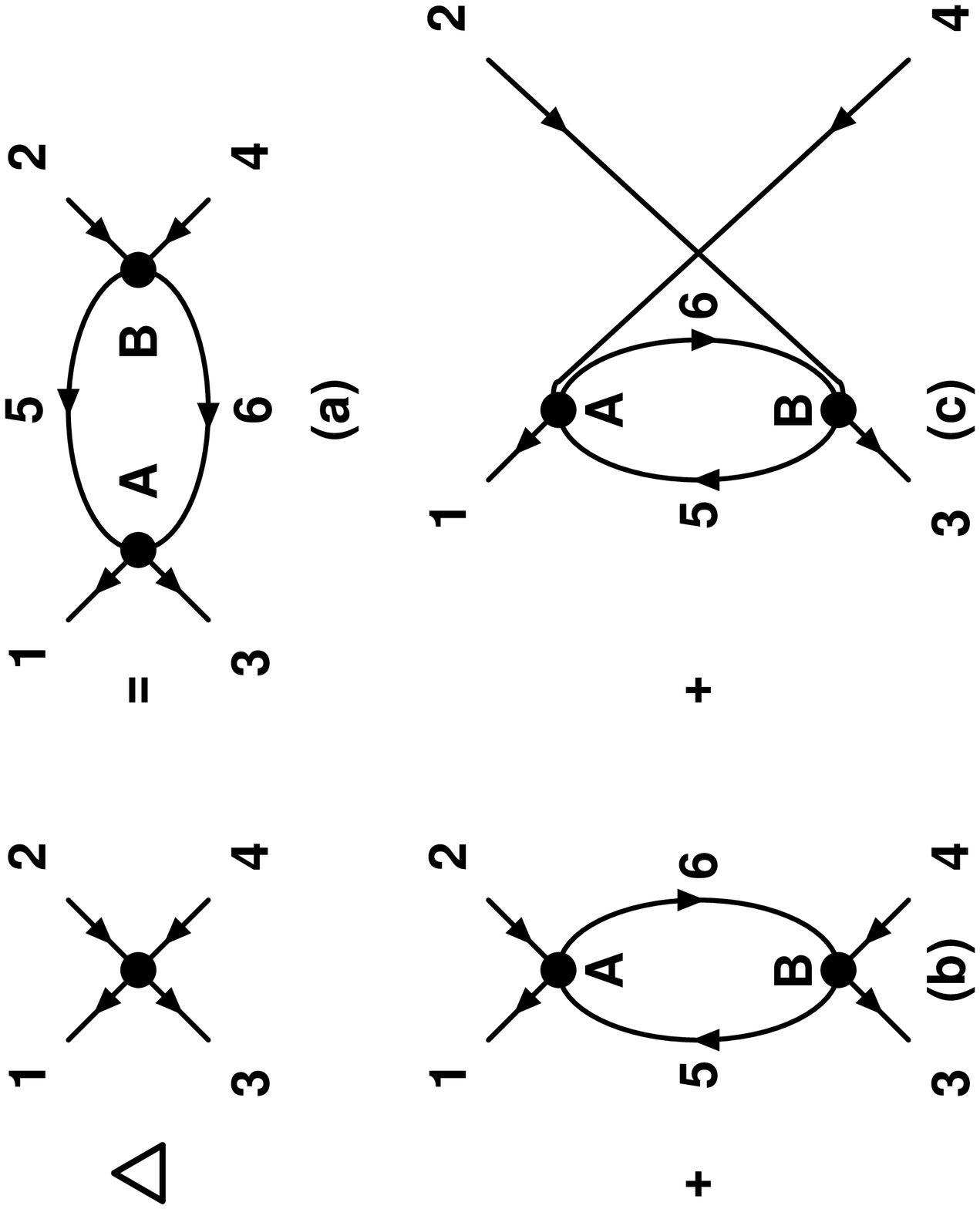}}}
\mycaption{ FIG. \figaa. One-loop diagrams contributing to  the 
  renormalization of the couplings $\ggg{}1342$}. 
 \end{figure}
}

\long\def\figb{
\begin{figure}[htb]
\centerline{
\psfig{width=9truecm,angle=-90,bbllx=50pt,bblly=100pt,bburx=560pt,bbury=750pt,file={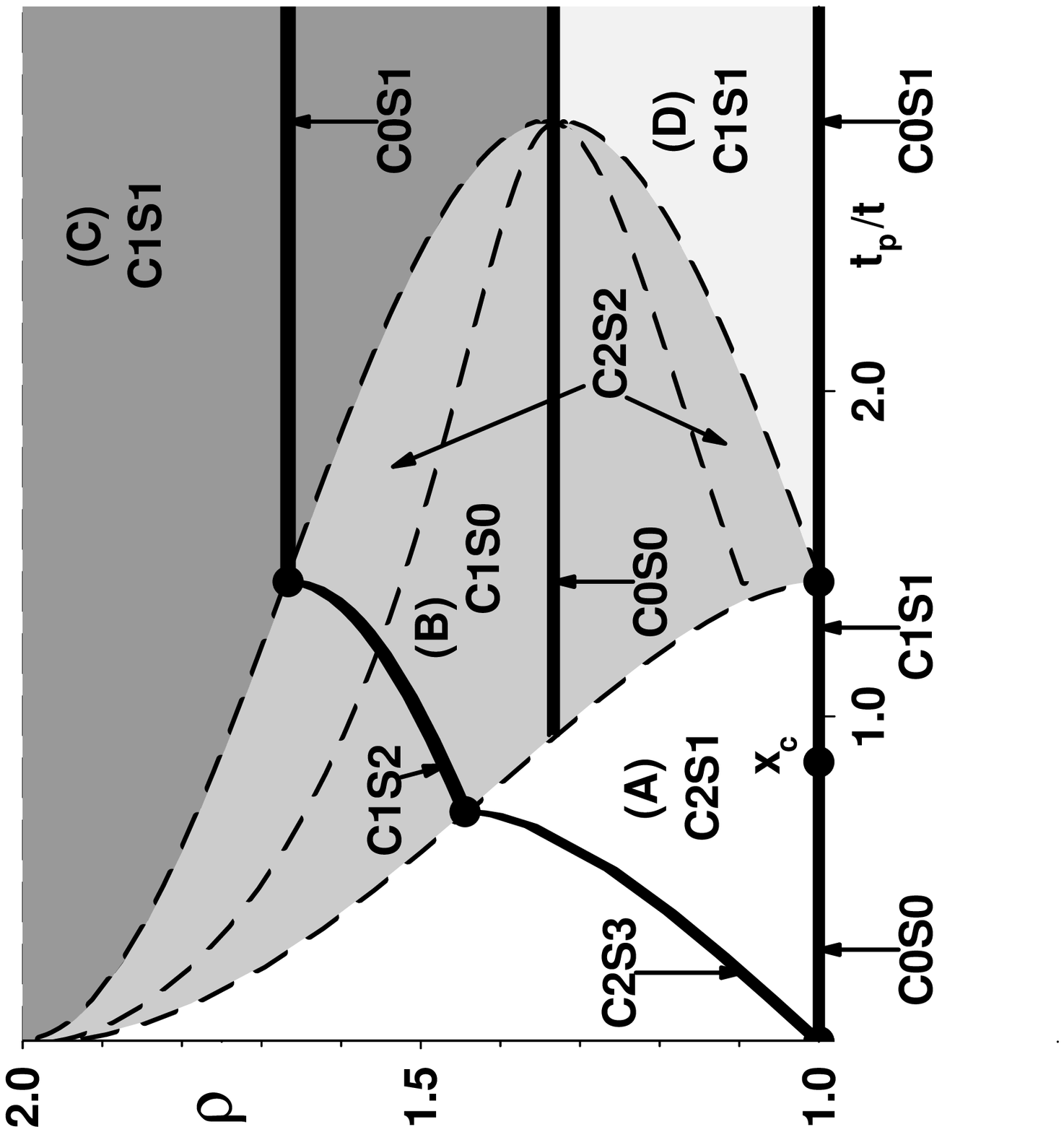}}}
\mycaption{ FIG. \figbb.
Phase diagram for the three-legs Hubbard ladder with open transverse boundary
conditions (OBC) as a function of transverse
hopping $t_{\pperp}>0$, and particle density $\rho\ge1$.
Thick lines represent regions ({\evi phase lines}) 
where a phase occurs
{\evi for a strict relation between the Fermi momenta} in the 
$U\to 0^+$ limit.
Phase boundaries are indicated by dashed lines for the extended regions 
and by dots for the phase lines.
Conventions for the labeling of the 
different phases are explained in the text.
\tagliac{ The four regions 
(A), (B), (C), and (D)  correspond to  different Fermi-surface
topologies (see text).}
}
 \end{figure}
}

\btwocols

The problem of 
coupled one-dimensional (1-D) systems of interacting fermions
(ladders)\cite{fabrizio,noack,balents,schulz} has recently 
received considerable attention  among
condensed-matter physicists 
due to the possible connections with
high-Tc superconductivity.
In addition, these systems present unusual properties whose analysis
can  give  valuable new insights into the
properties of low-dimensional quantum systems.
An interesting topic addressed in this field is the contrast between
the magnetic properties of ladders with even and odd numbers of legs,
in analogy with 1-D systems with integer and half-integer 
spin.\cite{noack,rice}

This  theoretical interest has been further
 prompted  
by the discovery of 
the Sr$_{n-1}$Cu$_{n+1}$O$_{2n}$ 
compounds,  whose lattice structure shows strips of 
 $(n+1)/2$-legs fermionic ladders.\cite{exp}
Different measures 
 on these  systems,
show that a spin gap 
(signaled by an exponential falloff of the
magnetic susceptibility as the temperature decreases)
  occurs
 in the even-legs compounds (at least at half filling), while
the odd-legs systems remain ungapped.

\tagliad{
A powerful  technique to study the low-energy properties of one- and 
quasi-1-D fermionic systems is bosonization.\cite{emery} 
 This method can also be
extended
to higher dimensions whenever one can restrict to  interactions that
are strongly
peaked about low values of the transferred momentum, thus
providing a good starting point to study systems for which
 ordinary Fermi-liquid behavior
cannot be a priori assumed.\cite{kopietz}
} 
In a strictly 1-D spin $1/2$ 
system the low-lying spin and charge excitations are completely
separated  and
the associated correlation functions show long-range oscillations with 
an anomalous power-law decay, whenever high-momentum-transferring processes
(backward and umklapp scattering) can be neglected, allowing for the
low-energy problem to 
be solved exactly.\cite{emery}
At   half filling, however, 
 the coupling associated with the umklapp process becomes {\evi relevant} 
 and opens a gap in the charge excitations  
causing the system to be an insulator (at $T=0$) 
for  arbitrary small  repulsion.\cite{solyom}
 This does not happen for the repulsive backward scattering
which remains {\evi irrelevant} 
leading to quasi-long-range-ordered
spin degrees of freedom.
The problem is thus to understand when a given 
 processes is important (relevant)
for the low-energy (or, equivalently, long-range) properties of the system.
This issue is best addressed  in the framework of the renormalization 
group (RG), whereby
one follows the evolution (flow)  of the couplings 
as the system is observed by larger and larger length scales.\cite{rg}
\tagliad{
 The couplings are then classified
into relevant or irrelevant depending on whether their size increases or
decreases along the flux.
For example, in the
 half-filled  system 
 the umklapp coupling 
 flows 
towards strong-coupling values, whereas a positive
 backward coupling flows to zero
 (is irrelevant) allowing 
the spin correlation function to remain quasi-long-range ordered
 (that is, it decays with a power law).}

The same question can be addressed for multi-legs systems, where
the analysis is complicated by 
the larger number of parameters.
 Extensive RG studies  have been carried out
  for two-legs ladders,\cite{fabrizio,balents,schulz} and for spin
  systems with an higher number of legs\cite{noack}.
The results of
 these calculations clearly show that 
the crossover from a 1-D
system to a system with many legs  
is far from being smooth at least
concerning 
 spin
excitations, which are exponentially 
suppressed in even-legs system, while they are
quasi-long-range ordered in systems with an odd number of legs.
Data from  the Sr$_{n-1}$Cu$_{n+1}$O$_{2n}$  ladder systems
agree with these results.\cite{exp}
A smooth behavior as a function of the number of legs 
should be then recovered when either even-legs or
odd-legs systems are considered separately. In this sense, a three-legs
system  is the first continuous extension of the
purely 1-D system.
Anomalous exponents for the uniform charge degrees of freedom should
instead continuously decrease with the number of legs, at least away
from half filling.\cite{schulz}

It is thus worthwhile to address the properties of an odd-legs system,
by extending the analysis also to the 
 {\evi charge} degrees of freedom and by
considering the behavior of the system upon doping.
In this paper, we study the phase
diagram of a three-legs spin $1/2$ fermionic ladder
by means of
 weak-coupling RG approach 
like the one  applied to one \cite{solyom} and
two coupled \cite{fabrizio,balents} one-dimensional systems, here
restricting to one-loop order.  
 We mainly analyze 
 the case of open boundary conditions in the  $y$
 direction (OBC),
 since it represents  the
experimental situation and it is free from frustration in the magnetic
interactions. 
 Our results yield indeed   ungapped  spin modes  
in a wide region about half filling, and at  half filling
 for an almost isotropic system.
 In the same region, the system is a conductor,
whereas a metal-to-insulator transition appears at half filling for
low values of $t_{\pperp}$.

Specifically, 
we consider a system of $3$ identical one-dimensional single-band models
(legs), each
 with band energy $\eps(k)$ ($|k|\leq\pi$), coupled via a single-particle
transverse hopping $t_{\pperp}$ and supplemented with
a generic short-range 
interaction.
For the sake of definiteness, we will specialize to the Hubbard model
with on-site repulsion $U$ and 
nearest-neighbor hopping $t$, although we will eventually allow for
extensions. 
 By diagonalizing the
single-particle term containing $t_{\pperp}$ and $\eps(k)$
 one obtains $3$ ``parallel''
bands with energies 
$ \eps(k,\al) = \eps(k) - T_{\pperp}(\al) $
where $\al = 1,\dots,3$ labels the $3$ bands, and 
 $ T_{\pperp}(\al)=
\sqrt2 \, t_{\pperp} (\al-2) $ 
or 
$ 2 \, t_{\pperp} \cos\frac{2 \pi (\al-1)}{3}$ 
for OBC and
 periodic boundary conditions in the $y$ direction (PBC), respectively.

Within an RG approach one first "integrates out" the
degrees of freedom 
far from the Fermi surface (consisting of 
6 points  when all bands cross the Fermi surface)
and restricts the range of the 
 momenta to regions up to a cutoff $\Lam$ about each
Fermi point,  such that, within each
of these disconnected regions, the dispersion can be
approximated with a linear function  and the interactions can be
taken as $k$-independent.
  The renormalization of the
parameters (Fermi momenta, Fermi velocities, and interactions) during
this ``preliminary'' integration can be neglected in the weak-coupling
limit and 
their bare values can be used as ``input'' for the
linearized system.\cite{balents}\tagliad{$^,$\cite{general}}

 The
generic interaction limited to  the restricted regions 
can thus be written
\beqn
 \sum_{ k,k',q \atop \g,\g'} \ \sum_{r_l,\al_l} && 
\ggg{\g\cdot\g'}{r_1,\al_1}{r_3,\al_3}{r_4,\al_4}{r_2,\al_2}\ 
\nonumber \\ && \times
\ccd{1}{k+q}{\g} \, \cc{2}{k}{\g} \,
 \ccd{3}{k'-q}{\g'} \, \cc{4}{k'}{\g'}
\nonumber \\ && \times
\delta_P\left(\sum_{l=1}^4 (-1)^l \ r_l\ \KF(\al_l)\right)\;
\label{gs}
\eeqn 
 where 
  $\cc{l}k{\g}$ ($\ccd{l}k{\g}$) is the destruction
 (creation) operator for
  a fermion with spin $\g$,
 momentum $k$ relative to the Fermi point
$r_l \cdot \KF(\al_l)$, and belonging to
 band $\al_l$. 
In \eqref{gs} the sum over momenta is restricted to 
$|k+q|, |k|, |k'-q|,  |k'|  \leq \Lam$, and
 $r_l=\pm 1$ indicates the branch of the Fermi surface where the
 particle resides. Conservation of crystal momentum is enforced by
 the function $\delta_P$.

The allowed values of the couplings $g$ are, of course, constrained
 by symmetries such as  conservation of crystal momentum in the 
longitudinal ($x$) 
direction,  spin conservation, 
parity under
reflection about the $x$ axis,
(in the PBC case, 
also transverse momentum conservation),
 time reversal and 
particle-exchange symmetry.
These symmetries
severely restrict the allowed couplings, especially if 
there is no particular relation between the Fermi momenta at the 
given value of the particle density.

 In the OBC case,
 the allowed couplings
can be classified like  for the two-legs system,\cite{varma}
where
 two band indices $\al_l$ in \eqref{gs} belong to one band and the
 other two to
  another  or to the same band,
alternatively with parallel or antiparallel spin combinations,
whereby the
$r_l$ are fixed by  $x$ momentum
conservation.\cite{details}
On the contrary,  for PBC the situation is more complicated  
since (a) some processes
\tagliad{
(a) processes in which two
bands are involved and in each band the particles belong to opposite
branches [the couplings $\Gamma_{AABB}$ of Ref.\onlinecite{varma}]
}
 are forbidden by    transverse momentum conservation, and (b)
the degeneracy of the two bands\tagliab{\cite{degeneracy}} 
$\al=2$ and $\al=3$
allows  for 
anomalous processes in which 
particles are scattered among all three bands.
In addition, more couplings are 
present at special particle densities 
both in the OBC (for example, at half filling) and in the PBC
case.\cite{balents} 

The RG flow is
 generated by reducing the momentum cutoff
$\Lam$ by infinitesimal steps
and by modifying (renormalizing) the parameters of the Hamiltonian in
such a way that the partition function remains fixed.\cite{rg}
At one-loop order,
  the differential
 renormalization of the  parameters is evaluated
by  integrating  the momenta of the internal  lines over an
infinitesimal shell  of width $\Delta \Lam $ at the cutoff
in the corresponding diagram.\tagliab{\cite{multi}$^,$}\cite{fabrizio} 
 The differential
renormalization 
$\Delta \ggg{\g\cdot\g'}{r_1,\al_1}{r_3,\al_3}{r_4,\al_4}{r_2,\al_2}$ 
 of a given coupling 
is simply obtained by
evaluating the ``bubble'' diagrams 
of Fig. \figaa with zero external momenta and wave vectors,
 whereby the internal wave vectors $q$ are restricted to the
 infinitesimal ``shell'' $\Lambda - \Delta \Lambda < |q| < \Lambda$. 
 The
  integration is  readily evaluated yielding from
 each diagram 
the
 contribution $
\eta\, \delta_{r_5+r_6}\, g_A \, g_B \, \frac1{2 \pi(|v_{5}|+|v_{6}|)} \,
\frac{\Delta \Lam}{\Lam} \,,$
where $v_{5}$ and $v_{6}$ are  the Fermi velocities of the internal lines
$5$ and $6$, respectively, and $\eta$ is a symmetry and sign
factor.\cite{details}  
The total contribution to the renormalization of a given couplings
corresponding to given external ``legs'' $1,2,3,4$ of the diagram 
is then obtained by summing
over all possible sets of 
labels ($\al_l=1,\dots,3$, $r_l=\pm1$, and $\g=\pm1$)
 for the internal lines $5$ and $6$  {\evi corresponding
 to allowed couplings $g_A$ and $g_B$}  for the 
 points A and B.\cite{details}
  In particular, it follows as usually  that
couplings with $r_1=r_2=r_3=r_4$ do not renormalize at the lowest
order.\cite{fabrizio,balents,varma}
The
effect of the renormalization of Fermi velocities and momenta
 can be neglected at this order.

\ifiga

After carrying out the  burdensome but
straightforward summation for  
 the different couplings,  by means of
a computer algorithm
for symbolic calculations,
we end up with a
set
of quadratic RG differential equations,
(40 equations at half filling, 27
for general density, in the OBC case,
 24 in the PBC case)
that can be   written:
\beq
 \frac {d\ g_i}{d \ \tau} =
A^i_{j,l} \, g_j \, g_l  \;, 
\label{eqs}
\eeq
 where $\tau=- \ln \Lam $,
 the indices $i,j$, and $l$ label the allowed
processes, and the matrix elements
 $ A^i_{j,l}$  depend
 on the Fermi velocities\tagliac{ (some elements,
related to  couplings allowed only at
special combinations of Fermi momenta, will be nonvanishing only at
the associated particle densities)}.

The precise expression for the system of differential equations \eqref{eqs}
would be too long to reproduce here,\cite{details}
 however, it can be readily solved
numerically by inserting initial values for the couplings depending on
the particular model.
The purpose here is to determine which couplings are most {\evi
  relevant} in the RG sense, that is, which ones of the couplings in
\eqref{gs} are important for the low-energy properties of the system.
This is achieved by looking at the behavior of their
RG flux.
It turns out that, depending on the initial conditions, some
of the couplings will diverge at a certain cutoff scale $\tau=\tau_0$ 
and some not (only
in few cases, all couplings converge).
In the weak-coupling limit the {\evi renormalized} Hamiltonian,
whereby only the relevant couplings are retained, can be constructed by
 fitting the asymptotic behavior of each coupling by the power law $
g_i(\tau)\approx \frac {g_{i 0}}{(\tau_0-\tau)}$, and by replacing
each $g_i$ by its associated
$g_{i0}$ in the Hamiltonian.\cite{balents} This procedure is {\evi
  controlled}  in the
weak-coupling limit since one can always choose an arbitrary small $U$
such that the asymptotic behavior occurs in the weak-coupling regime.
\tagliad{
 with a power law 
$ g_i(\tau)\approx \frac {g_{i
0}}{(\tau_0-\tau)^{\gamma_i}}$, 
 with $\gamma_i \le 1$,
and the others  will go to a finite value or to $0$.  
Notice that the invariance of Eqs. \eqref{eqs} under the simultaneous
 transformations
 $g\to g\eps$, and 
$\tau \to \tau/\eps$, implies 
$\tau_0 \to \tau_0/\eps$, and
  $ g_{i 0} \to \eps^{1-\gamma_i} g_{i0}$,
 For this reason,
 in the weak-coupling limit
(which corresponds to $\eps$ arbitrarily small)
  we can restrict to a renormalized Hamiltonian 
where only  the relevant
couplings  associated with a
nonvanishing
 $g_{i0}$ with $\gamma_i=1$ are retained.
The values of such $ g_{i 0}$ 
can be readily obtained by a  fit
of  Eqs. \eqref{eqs} with the asymptotic 
expressions.\cite{balents}\tagliad{$^{,}$\cite{system}}  
}

In order to understand the low-energy physics that follows from the
renormalized Hamiltonian, we
 write it in terms of bosonic fields,
\cite{balents} by expressing  the fermionic
destruction operators in the real-space representation
as\cite{emery}\tagliad{$^{,}$\cite{fourier}}
\beqn
 c_{r,\al,\g}(x) \approx && C \sqrt{\Lam} 
\exp[i (r \KF(\al)\, x -
  (r \phi_{\ch,\al}(x) - \theta_{\ch,\al}(x))
\nonumber \\ &&
+ \g (r
\phi_{\sp,\al}(x) - \theta_{\sp,\al}(x)))] \;,
\label{fermion}
\eeqn
 where $C$ is a constant of  order unity, and $\phi_{p,\al}(x) \ ,\ 
\theta_{p,\al}(x)$ are dual canonical bosonic fields associated with
the 
charge ($p=\ch$) or spin ($p=\sp$) modes on band $\al$.

\ifigb

The resulting bosonic Hamiltonian consists of two kinds of couplings.
Couplings of the {\evi density-density} type (containing operators of the type 
$ \, c^{\dag}_{r,\al,\g}(x) c_{r,\al,\g}(x) c^{\dag}_{r',\al',\g'}(y)
c_{r',\al',\g'}(y)$) 
contribute to the quadratic ungapped part of the 
Hamiltonian by  mixing  the different band modes, and renormalizing
the Fermi velocities. The model restricted to density-density type
couplings  {\evi conserves spin and charge at each point of the Fermi
  surface and is the analogue of a Fermi liquid in quasi-one
  dimension}, it  is exactly solvable and
shows spin-charge separation as well as a power-law behavior of
correlation functions.  However, the rest of
the couplings (which  will be here 
referred to  as boson-interaction couplings) 
add to the Hamiltonian complicated ``cosinus''-like  terms
 that cannot be treated exactly.  The standard way
to deal with these terms, whenever they turn out to be relevant,
is to expand the bosonic fields about one of
the (equivalent) minima of the interacting 
Hamiltonian thus yielding a mass gap for
the associated bosonic field.\cite{balents,mass} 

We first analyze the region
 where all three bands cross the Fermi
surface [region (A) of Fig. \figbb] 
and  no special combinations of Fermi momenta occur.
As it is clear from the discussion above, we are interested in
determining when some of the three charge and three spin modes
(one for each band)
acquire a gap due to the relevant processes. For this reason, we only need
 to consider the relevant couplings 
which are of the boson-interaction type.
In this case, one has two ``backward'' couplings $g_b(i)$  reversing the spin
in the two opposite branches of each of the two bands $i=1$ and $3$, and two
processes $g_{fb}(1,3)$  transferring particles  that move in opposite
directions from band $1$ to $3$ and vice-versa (with different spin
symmetry).  Band $2$ turns out to be
asymptotically free.
The associated boson-interaction term, obtained by using Eq. \eqref{fermion}
 reads:
\beqn
&&   g_b(1) \, \cos (4\,\phi_{\sp, 1})\, +  
         g_b(3) \, \cos (4\,\phi_{\sp, 3})\,
\nonumber \\ &&
+ 2 \, g_{fb}(1,3) \, 
\sin(2 \phi_{\sp,1})\, \sin(2 \phi_{\sp,3}) 
\cos
 [2(\theta_{\ch,1}-\theta_{\ch,3})] \,.
\label{bint}
\eeqn
  Expansion about the 
minimum at $\phi_{\sp,\al}=\pi/4\quad$ ($\al=1,3$)
and $\theta_{\ch,-}\equiv\theta_{\ch,1}-\theta_{\ch,3}=\pi/2$, yields
a mass
gap
for the two spin modes $\phi_{\sp, 1}$ and $\phi_{\sp, 3}$, and for the
odd charge mode $\theta_{\ch,-}$.
Using the  same
notation as in Ref.\onlinecite{balents}, whereby  a phase with
 $n$ gapless charge and $m$ gapless spin modes
is referred to as  C$n$S$m$, we have here a C2S1 phase (see Fig. \figbb).
 Note that, unlike the
two-legs system, 
where a C$1$S$0$ phase occurs,\cite{balents}
there are  {\evi ungapped spin modes} about half
filling, which implies, for example, that the spin correlation
function does not decay exponentially. 

The situation is quite different for the PBC system. In this case,
several boson-interaction couplings can become relevant, including some of
the processes that involve particles in the three different bands.
Two regions  can be distinguished:
 When the Fermi velocity $v_{F1}$ of the first band is larger than
the velocities of the other two (degenerate) bands $v_{F23}$, 
the relevant couplings produce
a C2S1 phase 
\tagliad{
 two gapped spin modes  ($\phi_{\sp, 2}+\phi_{\sp, 3}$, and $\theta_{\sp,
2}-\theta_{\sp, 3}$) and one gapped charge mode $\phi_{\ch, 2}-\phi_{\ch,
3}$ yielding a C2S1 phase 
}
similarly to  OBC.
On the other hand, when $v_{F1}<v_{F23}$
\tagliad{
, which corresponds to 
$\mu>-t_{\pperp}/2$ ($t_{\pperp}>0$),
 one obtains additional gaps for the
spin mode 
$\phi_{\sp, 1}$, and for the charge mode
 $\theta_{\ch, 1}-\frac12 \theta_{\ch, 2}-\frac12 \theta_{\ch,3}$
}
only the uniform charge mode is ungapped
 yielding a C1S0  phase.
Notice  that  this last region includes half
filling,\tagliac{\cite{hf}} where a  spin-liquid phase  is expected,
due to the frustration of  the periodic system.

At half filling,
 a number of additional couplings are allowed
 for the OBC system.
Due to the relation 
 $\KF(1) + \KF(3) = 2 \KF(2)=\pi$, one can have
um\-klapp processes with several
combinations of Fermi momenta all equal to $2\pi$.
\tagliad{
, namely, 
$4 \KF(2)$, $ \KF(1) +2 \KF(2) + \KF(3)$, and $2
\KF(1) + 2 \KF(3)$. 
}
It follows that, in addition to the usual
 um\-klapp term for band $\al=2$ 
 one has combined um\-klapp
processes for bands $\al=1$ and $\al=3$
as well as processes that scatter particles from and to
 the three different bands.
 The behavior of the RG flow for these couplings  turns out to be 
different depending on the value of $t_{\pperp}$.  For $t_{\pperp}/t>x_c
\approx 0.86$ the renormalized Hamiltonian is restricted to 
 the combined $1-3$ umklapp terms (equal to $g_{fb}(1,3)$) and  to
the interaction terms
\eqref{bint}
 (and to unimportant density-density terms).
Therefore,
in addition to \eqref{bint} one  has a bosonic interaction of the form
$
g_{fb}(1,3)  \, \cos[2(\phi_{\ch,1}+\phi_{\ch,3})] 
[\cos(2\theta_{\ch,-})+2\cos(2\phi_{\sp,1})
 \cos(2\phi_{\sp,3})]
$
 which leads to an additional gap in the
$\phi_{\ch,1}+\phi_{\ch,3}$ charge mode and to  a C1S1 phase.
Remarkably, the umklapp process on band $2$ is suppressed by the
interaction with bands $1$ and $3$ and is thus
irrelevant in this
region.  The system is thus a
 {\evi conductor} 
(the current being carried by the
band $\al=2$)
with an {\evi ungapped
spin mode}.\cite{isol}

 For $t_{\pperp}/t < x_c$ a number of
additional couplings, including some of the 
processes that involve
  particles in the three different bands become
 relevant.
The analysis of the complicated
boson-interaction Hamiltonian\cite{details} 
leads to a C0S0 phase with {\evi completely gapped 
charge and spin modes}  (like
for the two-legs system at half filling\cite{balents}).
 One therefore has a {\evi metal-to
insulator transition} for decreasing $t_{\pperp}$ at $t_{\pperp}/t =
x_c$.  Notice that,  regarding the  spin modes,
 our results are consistent with 
the experiments (which probe an essentially isotropic system, thus
$t_{\pperp}/t \approx 1$).\cite{exp,isol}
  It would be
interesting to verify whether in anisotropic spin ladders
($t_{\pperp}< t $) a spin gap is present at half filling 
or whether our result is due to the
 weak-coupling condition.\tagliad{\cite{tp0}} 

The half-filling condition for band $1$
is fulfilled on the upper 
thick line of Fig. \figbb. In this region
\tagliad{
(we consider here only the
OBC case\tagliac{ whereby a symmetry under the exchange of the two extremal
bands $\al=1$ and $\al=3$ occurs}).
In this case, as for  two legs,  we obtain that only the
associated um\-klapp coupling 
diverges and suppresses all other divergences. The
system has thus only
}
a single gapped charge mode occurs.\cite{balents}
We shall neglect 
  other special cases which can arise due to 
particular combinations of
 Fermi momenta or when a band edge touches the
 Fermi level.\cite{balents} 
\tagliac{Like half filling,
 they  represents {\evi phase lines} in the weak-coupling phase diagram
that can, possibly broaden and shift when the interaction is increased.}

In region (B) of the phase diagram  
 only two bands cross the
Fermi surface. The analysis of this region  is 
identical to that for the two-legs
ladder\cite{balents} yielding  the same set of RG equations and
similar phases. 
\tagliad{
However, 
the different initial conditions for the couplings yield a  shift of the
phase boundaries and  a reduction of the size of
 the internal C1S0 region on behalf of  the C2S2 peripheral phase 
(the intermediate C2S1 phase disappears). }
Finally, in the remaining regions (C) and (D) 
 only one band 
is involved 
\tagliad{($\al=1$ and $\al=2$, respectively)} 
and one recovers
the results of the purely one-dimensional system.\cite{solyom}

In order to check the dependence on the detailed
features of the particular model,  we
have  tested our results by 
 including, alternatively,
 a  nearest-neighbor  $U_{nn}$ and a transverse
$U_{\pperp}$
interaction.
While values of  $U_{nn}/U$ up to $0.5$ essentially don't modify the
phase diagram, it turns out that already with 
 $U_{\pperp}/U \cmag 0.2 $ charge and spin excitations tend to be suppressed
and additional gaps arise especially in the neighborhood of the phase
boundaries, both at half filling and at finite doping.

In conclusion, we have investigated the weak-coupling phase diagram of
three coupled one-dimensional electronic systems modeled by a
Hubbard Hamiltonian (allowing eventually for some extensions) 
with open boundary conditions in the transverse 
direction. 
 Consistently with  experiments on 
Sr$_2$Cu$_3$O$_{5}$,
spin modes are ungapped in a large  region at
finite doping and at half filling for $t_{\pperp}/t$ 
greater than a critical value $x_c$. 
In the same region, the system is a conductor for $U$ small enough and
 a Mott transition
occurs at half filling  
by decreasing  $t_{\pperp}$.
We have also analyzed the
 system with periodic transverse boundary conditions which 
shows, in contrast, 
a  spin gap in a large region at and close to half
filling.

As it was clear
 for the two-legs case,\cite{balents}
the interest 
of this
 analysis applied to a quasi-one-dimensional
system relies upon 
the  non-trivial
 phase diagram
which arises already at weak coupling. Here, the system is {\evi
  strongly correlated even at infinitesimal $U$}.
By increasing $U$, one should expect
a smooth
deformation of the phase boundaries and a widening of the isolated
phase lines in Fig. \figbb,
 such that the general features of the phase diagram should not
change dramatically  for finite $U$ (up to  a certain value\cite{isol}).

We thank G. C. Strinati, B. Brendel, J. Voit, H. J. Schulz, and W. Hanke 
for  stimulating discussions and precious suggestions.
 The author acknowledges research support from the ECC
under  the network program N. ERBCHRX-CT940438.


\efiga
\efigb

\etwocols

\end{document}